\documentclass[12pt]{iopart}

\usepackage{iopams}

\expandafter\let\csname equation*\endcsname\relax

\expandafter\let\csname endequation*\endcsname\relax

\usepackage{amsmath}
\usepackage{amsfonts}
\usepackage{amssymb}
\usepackage{hyperref}
\usepackage{graphicx}
\usepackage{color}
\usepackage{float}
\usepackage{wrapfig}
\usepackage{soul}
\usepackage{epsfig}
\usepackage{cite}
\begin{document}

\title[]{Complexity-like properties and parameter asymptotics of $\mathfrak{L}_{q}$-norms of Laguerre and Gegenbauer polynomials.}

\author{Jes\'us S. Dehesa}
\address{Instituto Carlos I de F\'{\i}sica Te\'orica y Computacional, Universidad de Granada, Granada 18071, Spain}
\address{Departamento de F\'{\i}sica At\'{o}mica, Molecular y Nuclear, Universidad de Granada, Granada 18071, Spain}
\vspace{10pt}
\ead{dehesa@ugr.es}
\author{Nahual Sobrino}
\address{Donostia International Physics Center, Paseo Manuel de Lardizabal 4, E-20018 San Sebasti\'an, Spain}
\address{Nano-Bio Spectroscopy Group and European Theoretical Spectroscopy Facility (ETSF), Departamento de Pol\'imeros y Materiales Avanzados: F\'isica, Qu\'imica y Tecnolog\'ia, Universidad del Pa\'is Vasco UPV/EHU, Avenida de Tolosa 72, E-20018 San Sebasti\'an, Spain}
\vspace{10pt}\ead{nahualcsc@dipc.org}

\begin{abstract}
  The main monotonic statistical complexity-like measures of the Rakhmanov's probability density associated to the hypergeometric orthogonal polynomials (HOPs) in a real continuous variable, each of them quantifying two configurational facets of spreading, are examined in this work beyond the Cram\'er-Rao one. The Fisher-Shannon and LMC (L\'opez-Ruiz-Mancini-Calvet) complexity measures, which have two entropic components, are analytically expressed in terms of the degree and the orthogonality weight's parameter(s) of the polynomials. The degree and parameter asymptotics of these two-fold spreading measures are shown for the parameter-dependent  families of HOPs of Laguerre and Gegenbauer types. This is done by using the asymptotics of the R\'enyi and Shannon entropies, which are closely connected to the $\mathfrak{L}_{q}$-norms of these polynomials, when the weight's parameter tends towards infinity. The degree and parameter asymptotics of these Laguerre and Gegenbauer algebraic norms control the radial and angular charge and momentum distributions of numerous relevant multidimensional physical systems with a spherically-symmetric quantum-mechanical potential in the high-energy (Rydberg) and  high-dimensional (quasi-classical) states, respectively. This is because the corresponding states' wavefunctions are expressed by means of the Laguerre and  Gegenbauer polynomials in both position and momentum spaces. 
\end{abstract}

\section{Introduction}

The quantification of the spreading of the hypergeometric orthogonal polynomials (HOPs) along the support interval $\Lambda \subseteq{\mathbb{R}}$ is interesting \textit{per se} in the theory of special functions and approximation theory, and because of their numerous applications in quantum mechanics and mathematical physics \cite{Nikiforov1988,Temme1996,Andrews1999,Ismail2005,Olver2010,Koekoek2010}. A relevant reason for the latter is that the HOPs control the physical solutions  of the non-relativistic and relativistic wave equations of a great deal of relevant quantum systems (oscillator-like systems, hydrogenic atoms,...)  \cite{Nikiforov1988,Bagrov1990,Cooper2001}. Here we study the spreading measures of the real HOPs, $\{p_n(x)\}$, orthogonal with respect to the weight function $h(x)$ on the support  interval $\Lambda$. These quantities are defined by the corresponding measures of the normalized-to-unity Rakhmanov's probability density $\rho_n(x) =  p_n^2(x)\,h(x)$. This density function governs the ($n\to+\infty$)-asymptotics of the ratio of two polynomials with consecutive orders \cite{Rakhmanov1977}, and describes the quantum-mechanical probability density of the bound stationary states of a great deal of quantum systems in one and many dimensions \cite{Nikiforov1988,Yanez1994,Dehesa2001,Dong2011,Brandon2013}. Indeed, it happens e.g. that the wavefunctions for the bound states of a large family of non-relativistic quantum-mechanical potentials are controlled by the three canonical HOPs families of Hermite $H_n(x)$, Laguerre $L_n^{(\alpha)}(x)$ and Jacobi $P_n^{(\alpha,\beta)}(x)$ types \cite{Nikiforov1988,Olver2010}. So, the associated Rakhmanov density $\rho_n(x)$ may be often interpreted as the charge and/or the matter density of single-particle quantum systems. Consequently, the spreading measures of the HOPs characterize different fundamental and/or experimentally measurable properties of physical and chemical systems. \\ 

Beyond the dispersion measures (the standard deviation and its extensions, the ordinary and central moments), the spreading measures of a given probability density have an entropy-like origin. Contrary to the dispersion ones, the entropic measures do not depend on any specific point of the density's support, so that they quantify spreading facets qualitatively different from the ones given by the dispersion measures. Then, the entropic measures quantify the different facets of the  extent of the density along its support in a much more appropriate manner.\\

The entropy-like measures, each quantifying a single spreading facet, are of local (Fisher information) or global (R\'enyi and Shannon entropies) character depending whether they are very sensitive to the fluctuations of the density or not, respectively. The Fisher information $F(\rho)$, which is the most familiar and relevant local entropic measure \cite{Fisher1925,Frieden2004}, is a functional of the derivative of the density $\rho(x)$. Then, it controls the localization of the density around its nodes, appropriately grasping the oscillatory nature of the density. This allows it to characterize a great diversity of scientific phenomena which are closely connected to the kinetic and Weizs\"acker energies \cite{Sears1980,Parr1989, Romera1994}  of the quantum systems. The R\'enyi entropies $R_{q}[\rho],\, q \neq 1,\,$\cite{Renyi1961,Renyi1970}, which depend on a real parameter $q$, and its limiting case $q\rightarrow 1$, the Shannon entropy $S[\rho]$ \cite{Shannon1948,Cover1991}, are the most important global spreading measures. They are $q$-power functionals of the density, closely related to the algebraic $\mathfrak{L}_{q}$-norms of the involved HOPs. Then, they can describe many quantities of great scientific and technological interest, such as e.g. the thermodynamical entropy in the case of a thermal ensemble and the disequilibrium when $q \rightarrow 1$  and $2$, respectively; moreover, they are the basic variables of the classical and quantum information theories \cite{Cover1991,Nielsen2000,Bruss2019}.\\

The knowledge of the entropic measures of the HOPs has been recently reviewed \cite{Aptekarev2010,Dehesa2021}. Therein, the analytical expressions for the Fisher information and the R\'enyi and Shannon entropies are given for the three canonical families of the real HOPs at all $n$. The ones for the global entropies are not handy in the sense that they \textit{only} provide algorithmic expressions to compute them in a symbolic way because they require the calculation of the combinatorial Bessel polynomials evaluated at the HOP expansion coefficients or some multivariate hypergeometric functions evaluated at unity for the R\'enyi cases, and the calculation of the logarithmic potential of the HOPs evaluated at their zeros for the Shannon case. For the most complicated situations (i.e., when the polynomial degree $n$ is high), however, the degree asymptotics ($n\rightarrow \infty$) of the R\'enyi and Shannon entropies and the related weighted $\mathfrak{L}_{q}$-norms of the HOPs allows one to obtain simple, transparent and compact expressions.\\

In this work we consider the monotonic complexity-like measures of the HOPs \cite{Rudnicki2016}, which are quantities composed by two or more entropic factors, so that they can simultaneously quantify two or more different configurational facets of the spread of the HOPs along the orthogonality interval. The idea in mind is to quantify the simplicity/complexity of the  HOPs as simpler and better as possible by means of a single measure. Three measures have been recently proposed \cite{Dehesa2015}, the Cram\'er-Rao, Fisher-Shannon and LMC (L\'opez-Ruiz-Mancini-Calvet)
complexities, which were originally introduced in a quantum-physical context (see e.g. the reviews \cite{Angulo2011} and \cite{Martin2006,Dehesa2009,Dehesa2011,Molina2012,Lopez2009,Lopez2013}). Up until now, however, it is only known \cite{Dehesa2015} the explicit expression of the Cram\'er-Rao complexity for the the three canonical families of HOPs at all $n$ and the asymptotical values for the three HOPs families when $n\rightarrow \infty$, assuming the weight function's parameters to be fixed.\\

 The main goal of this paper is to find the Fisher-Shannon and LMC complexities for the Laguerre and Gegenbauer orthonormal polynomials with an arbitrary fixed degree when the weight function's parameter tends to infinity. This issue is theoretically relevant in the theory of special functions, and because of its direct applications to compute the physical entropies of quasi-classical or high dimensional states of multidimensional systems in quantum physics and quantum tecnologies \cite{Puertas2017b,Puertas2017,Dehesa2019,Dehesa2020}; the latter is because the radial and spherical components of the state's wavefunctions are controlled by Laguerre and Gegenbauer polynomials with a parameter which depends on the space dimensionality of the systems, respectively. The Gegenbauer polynomials are well known \cite{Nikiforov1988,Olver2010} to be a particular family of the Jacobi polynomials. We will use the recent methodology of Temme et al. \cite{Temme2017,Puertas2017} which is based on the weight-function's parameter asymptotics of  the (unweighted) $\mathfrak{L}_{q}$-norms of the HOPs with a fixed degree. It is worth remarking that we do not consider here the complexity-like measures of polynomials with varying weights (i.e., polynomials whose weight-function's parameter does depend on the polynomial degree), which are also of great mathematical and physical interest \cite{Buyarov1999,Kuijlaars2004,Levin2018}. 
    \\

 This paper has the following structure. In Section \ref{density}  we describe the basic monotonic complexity measures with two entropic components of the Rakhmanov probability density for the HOPs.  In Sections \ref{FisherShannonLag} and \ref{LMCLag} we obtain the asymptotic behavior for the Fisher-Shannon and LMC complexities of the Laguerre polynomials when ($n\rightarrow \infty; \mbox{fixed}\, \alpha$) and when ($\alpha\rightarrow \infty; \mbox{fixed}\, n$) in a simple, compact form. In Sections \ref{FisherShannonGeg} and \ref{LMCGeg} we find the corresponding issue for the Gegenbauer polynomials. Finally, some concluding remarks are pointed out and a number of open related issues.

% Other global spreading facets of the position density $\rho(\vec r)$ are given by the monoparametric R\'enyi entropies \cite{renyi_61,leonenko}, $R_q[\rho],$
%\begin{equation}
%\label{REN}
%R_{q}[\rho]:=\frac{1}{1-q}\, \log \int_{\Delta}\rho(\vec{r})^{q}d\vec r, \quad q > 0, \quad q \not= 1.
%\end{equation}
%Note that these quantities include the Shannon entropy, the disequilibrium and the Tsallis
%entropies $T_{q}[\rho] =  \frac{1}{q-1} (1-\int_{\mathbb{R}^3} [\rho(\vec{r})]^{q})$,
%since $S[\rho] = \lim_{q\rightarrow 1} R_{q}[\rho]$, $\mathcal{D}[\rho] = \exp(-R_{2}[\rho])$ and
%$T_{q}[\rho] = \frac{1}{1-q}[e^{(1-q)R_{q}[\rho]}-1].$

\section{Complexity measures of Rakhmanov's density of HOPs}
\label{density}

In this Section we briefly describe the three basic complexity measures of the  HOPs $\{p_n(x)\}$ orthogonal with respect to the weight function $h(x)$ on the interval $\Lambda \subseteq{\mathbb{R}}$; namely, the Cram\'er-Rao, Fisher-Shannon and L\'opez-Ruiz-Mancini-Calbet (LMC) measures. They are defined \cite{Dehesa2015} by the corresponding complexity measures of the associated Rakhmanov's probability density 
\begin{equation}\label{rakhmanov}
	\rho_n(x) =  p_n^2(x)\,h(x),
\end{equation}
where the polynomials $\{p_n(x)\}$ fulfill the orthogonality condition \cite{Nikiforov1988,Olver2010}
\begin{equation}
\int_{\Lambda} p_n(x) p_m(x)\, h(x) dx = \kappa_n\, \delta_{n,m},\qquad \mathrm{deg}\, p_n=n
\label{eq:orthogonality_relation}
\end{equation}
and the weight function $h^{p}(x)$ on the support $(a,b)$ and the normalization constant $\kappa_n^{p}$ of the HOPs $\{p_n(x)\}$  considered in this work are given in Table \ref{Table_HOPS}. Note that $\kappa_n^{p}=1$ for the orthonormal polynomials $\hat{p}_n(x)$ of Hermite $\hat{H}_n(x)$, Laguerre $\hat{L}_n^{(\alpha)}(x)$, Jacobi $\hat{P}_n^{(\alpha,\beta)}(x)$ and Gegenbauer $\hat{C}_n^{(\lambda)}(x)$ types; so that the relation between the orthogonal and orthonormal HOP's is $p_{n}(x)= \hat{p}_{n}(x)\,(\kappa_n^{p})^{\frac{1}{2}}$.
\begin{table}[h!]
\centering
\begin{tabular}{|l|l|l|l|l|}
\hline
$p_{n}(x)$ & (a,b) & weight function $h^{p}(x)$& normalization $\kappa_{n}^{p}$ & constraints \\ \hline
$H_n(x)$ & $(-\infty,\infty)$ & $e^{-x^2}$ & $ \sqrt{\pi}\, n!\,\, 2^n$  &  \\ \hline
$L_n^{(\alpha)}(x)$ & $(0,\infty)$ & $x^\alpha e^{-x}$ & $\ \Gamma(n+\alpha+1)/n!$ & $\alpha>-1$ \\ \hline
$P_n^{(\alpha,\beta)}(x)$ & $(-1,1)$ & $(1-x)^\alpha (1+x)^\beta$ & $\frac{2^{\alpha+\beta+1}\Gamma(\alpha+n+1)
\Gamma(\beta+n+1)}{n!(\alpha+\beta+2n+1)\Gamma(\alpha+\beta+n+1)}$ & $\alpha,\beta>-1$ \\ \hline
$C_n^{(\lambda)}(x)$ & $(-1,1)$ & $(1-x^2)^{\lambda-\frac{1}{2}}$ & $\frac{2^{1-2\lambda} \pi \Gamma(n+2\lambda)}
{\left[\Gamma(\lambda) \right]^2(n+\lambda)n!},$ & $\lambda> -\frac{1}{2},\,\lambda\ne 0$ \\ \hline
\end{tabular}
\caption{Some properties of the hypergeometric orthogonal polynomials considered in this work.}
\label{Table_HOPS}
\end{table}

The Cram\'er-Rao complexity of the polynomial $p_n(x)$ is given \cite{Dembo1991,Dehesa2006,Antolin2009} by 
\begin{equation}
      \label{cramerrao}
      \mathcal{C}_{CR}[p_n]=F[p_n] \times V[p_n],
  \end{equation}
where $F[p_n]$ and $V[p_n]$ are the Fisher information \cite{Fisher1925,Frieden2004} and the variance of the Rakhmanov density $\rho_n(x)$ associated to $p_n(x)$, which are defined as
 \begin{equation*}
     F\left[p_n\right]=\int_{\Lambda} \frac{[\rho'_n(x)]^2}{\rho_n(x)}dx,\quad \mbox{and}\quad V[p_n]=\left(\Delta x\right)^2=\langle x^2 \rangle-\langle x \rangle^2,
  \end{equation*}
respectively, with the expectation value $\langle x^k \rangle= \int_{\Lambda} x^k \rho_n(x) dx$ for $k=1,2$. Then, the Cram\'er-Rao complexity quantifies the gradient content (so, the pointwise concentration of the Rakhmanov probability over its support interval) of $\rho_n(x)$ jointly with the spreading of the probability around the centroid.\\

The Fisher-Shannon complexity of the polynomial $p_n(x)$ is given \cite{Angulo2008,Romera2004} by
\begin{equation}
     \label{fishershannon}
      \mathcal{C}_{FS}[p_n]=F[p_n] \times \frac{1}{2 \pi e} e^{2 S[p_n]}=\frac{1}{2 \pi e} F[p_n] \times \left(\mathcal{L}_{S}[p_n]\right)^2,
  \end{equation}
where the symbol $S[p_n]$ denotes the Shannon-like entropic functional of the polynomial $p_n(x)$,
\begin{equation}\label{shannonfun}
     S[p_n] = \lim_{q\rightarrow 1} R_q[p_n]=-\int_{\Lambda} \rho_n(x) \log \rho_n(x) dx,
  \end{equation}
  which is the limiting case $q\rightarrow 1$ of the R\'enyi entropy of $p_n(x)$  defined as
  \begin{equation}\label{renyinorm}
       R_q[p_n] = \frac{1}{1-q}\log W_q[p_n], \quad \mbox{being}\quad W_q[p_n] = \int_{\Lambda} [\rho_n(x)]^q dx
   \end{equation}
   the $q$th-order entropic moment or weighted $\mathfrak{L}_{q}$-norm of the associated Rakhmanov density (\ref{rakhmanov}). The symbol $\mathcal{L}_{S}[p_n]=e^{S[p_n]}$ denotes the Shannon entropic power or Shannon spreading length of the polynomial $p_n(x)$. Note that the Fisher-Shannon complexity $C_{FS}[p_n]$ estimates the gradient content of the Rakhmanov probability density $\rho_n(x)$ associated to the polynomial $p_n(x)$, together with its total extent along the support interval $\Lambda$ of the orthogonality weight function $h(x)$. In addition, we remark that, from (\ref{rakhmanov}) and (\ref{shannonfun}), one has that the Shannon-like entropic functional can be expressed as
   \begin{equation}\label{shannon2}
   	S[p_n] = E[p_n] + I[p_n],
   \end{equation}
   where the symbols $I[p_n]$ and $E[p_n]$ denote the integral functional 
   \begin{equation}
   	I[p_n] = -\int_{\Lambda}%_{0}^{\infty}
\left[p_n(x)\right]^2 h(x) \log h(x)\,dx
   \end{equation}
and the Shannon entropy of the polynomial $p_n(x)$, 
   \begin{equation}\label{E_p}
   	E[p_n] = -\int_{\Lambda}%_{0}^{\infty}
\left[p_n(x)\right]^2 h(x) \log \left[p_n(x)\right]^2\,dx	,
   \end{equation}
 respectively.
  Note that the Shannon entropy of the orthogonal and orthonormal polynomials are related by
  \begin{equation}\label{E_pn}
  		E[\hat{p}_n] = \frac{1}{\kappa_n} 	E[p_n] + \log \kappa_n,
  \end{equation}
 and the corresponding relation for the Shannon-like entropic functionals is
 \begin{equation}\label{shannon3}
   	S[\hat{p}_n] = E[\hat{p}_n] + I[\hat{p}_n]= \frac{1}{\kappa_n} 	S[p_n] + \log \kappa_n,
   \end{equation}
 because $I[\hat{p}_n]= \frac{1}{\kappa_n}\,I[p_n]$.\\
 
 The LMC complexity of the polynomial $p_n(x)$ is defined \cite{Catalan2002} as
  \begin{equation}
     \label{lmc}
     \mathcal{C}_{LMC}[p_n]=W_2[p_n] \times e^{S[p_n]} = W_2[p_n] \times \mathcal{L}_{S}[p_n],
  \end{equation}
  which quantifies the combined balance of the disequilibrium of the associated Rakhmanov density or deviation from uniformity (as given by the averaging density $<\rho>$ or second-order entropic moment $W_2[p_n]$, which is a measure of order), and its total extent (as given by the Shannon entropic power $\mathcal{L}_{S}[p_n]$, which is a measure of disorder). Note for mathematical convenience that the disequilibrium of the orthogonal and orthonormal polynomials are mutually related by
  \begin{equation}\label{omega2ortho}
  	W_2[\hat{p}_n] = \frac{1}{(\kappa_n)^2}\, W_2[p_n].
  \end{equation}
 %\textcolor{red}{Chequear si es $\kappa_n$ o bien $(\kappa_n)^2$}. 
 These three (dimensionless) complexity measures of the HOPs polynomial $p_n(x)$ turn out (a) to grasp the combined balance of two different configurational facets of the associated Rakhmanov density, (b) to be bounded from below by unity (when $\rho_n(x)$ is a continuous density in $\mathbb{R}$ in the Cram\'er-Rao and Fisher-Shannon cases, and for any $\rho_n(x)$ in the LMC case), (c) to be minimum for the two extreme (or least complex) distributions which correspond to perfect order (i.e. the extremely localized Dirac delta distribution) and maximum disorder (associated to a uniform or highly flat distribution), and (d) to fulfil invariance properties under replication, translation and scaling transformation \cite{Yamano2004a,Yamano2004b}.\\

Finally, the Cram\'er-Rao complexity $\mathcal{C}_{CR}[p_n]$ has been explicitly found at all $n$ \cite{Dehesa2015} for the three canonical  HOPs families $\{p_n(x)\}$ in an analytical compact form, basically because the variance and Fisher information of their associated Rakhmanov densities are expressed in an analytically handy way. Such is not the case for the weighted $\mathfrak{L}_{q}$ norm nor for the Shannon-like entropic functional $S[p_n]$, so that the Fisher-Shannon and LMC complexity-like measures have not yet been analytically determined for all $n$, but \textit{only} for very high $n$ in the Fisher-Shannon case; the latter is basically because of the strong degree asymptotics of Aptekarev et al \cite{Aptekarev1994,Assche1995,Aptekarev1996} for the Shannon entropy of  HOPs polynomials of Hermite \cite{Sanchez2010,Aptekarev2012}, Laguerre \cite{Sanchez2011} and Jacobi \cite{Guerrero2010} polynomials.\\

 In this work, we extend the asymptotical knowledge of the Fisher-Shannon and LMC complexity-like measures of Laguerre and Gegenbauer polynomials for very high $n$ (LMC) and very high weight-function parameter (Fisher-Shannon, LMC). This is done in the following sections by use of both the degree asymptotics mentioned above and the parameter asymptotics of Temme et al \cite{Temme2017,Puertas2017} for the weighted $\mathfrak{L}_{2}$-norm and the Shannon entropy of the Laguerre and Jacobi polynomials. Let us advance that the main results obtained in the next four sections are collected in Tables \ref{Table_L} and \ref{Table_G} for the Laguerre and Gegenbauer polynomials, respectively.
 
 \begin{table}[H]
\centering
\begin{tabular}{|c|c|c|}
\hline
Measure  of $\hat{L}_n^{(\alpha)}(x)$ & n$\to \infty$ & $\alpha\to \infty$ \\ \hline
$F[\hat{L}_n^{(\alpha)}]$ & $\begin{array}{cc}4n & \alpha=0 \\ \frac{2\alpha}{\alpha^2-1}n & \alpha>1 \end{array}$ & $\frac{2n+1}{\alpha}$
 \\ \hline
$\mathcal{L}_{S}[\hat{L}_n^{(\alpha)}]$ & $\frac{2\pi}{e}n$ & $\frac{\sqrt{2\pi\alpha}}{n!}e^{n+\frac{1}{2}}$ \\ \hline
$W_{2}[\hat{L}_n^{(\alpha)}]$ & $\frac{\log n}{\pi^{2}n}$& $\alpha^{2n}\frac{1}{2(n!)^{2}\sqrt{\pi \alpha}}$  \\ \hline
$\mathcal{C}_{FS}[\hat{L}_n^{(\alpha)}]$ & $\begin{array}{ll}
      \left(\frac{8 \pi}{e^3}\right) n^{3} & \alpha=0 \\
      \frac{\alpha}{\alpha^2-1} \left(\frac{4\pi}{e^3}\right) n^{3} & \alpha>1
      \end{array}$
 & $ \frac{2n+1}{(n!)^2}e^{2n}$ \\ \hline
$\mathcal{C}_{LMC}[\hat{L}_n^{(\alpha)}]$ & $\frac{2}{\pi\,e}\,\log n$ & $\alpha^{2n}\left(\frac{e^{n+1/2}}{2^{\frac{1}{2}}(n!)^{3}}\right)$ \\ \hline
\end{tabular}
\caption{First order asymptotics for the entropy-like ($F, \mathcal{L}_{S}, W_2$) and complexity-like ($\mathcal{C}_{FS},\mathcal{C}_{LMC}$)   measures of the orthonormal Laguerre polynomials $\hat{L}_n^{(\alpha)}(x)$, $\alpha>-1$, when $n\to \infty$ and $\alpha\to \infty$. }
\label{Table_L}
\end{table}

\begin{table}[H]
\centering
\begin{tabular}{|c|c|c|}
\hline
Measure of $\hat{C}_n^{(\lambda)}(x)$ & n$\to \infty$ & $\lambda\to \infty$ \\ \hline
$F[\hat{C}_n^{(\lambda)}]$ & $\begin{array}{ll}
4\,n^3 & \lambda = \frac{1}{2} \\
\frac{2\lambda - 1}{\lambda^2-\lambda-\frac{3}{4}}\,n^3 
&\lambda > \frac{3}{2} \\
\infty & {\rm otherwise}
\end{array} $ & $(4 n+2)\lambda$ \\ \hline
$\mathcal{L}_{S}[\hat{C}_n^{(\lambda)}]$ & $ \frac{\pi\, 2^{1-2\lambda}}{e}$ & $\frac{(2\lambda)^{2n}}{(n!)^{2}}$  \\ \hline
$W_{2}[\hat{C}_n^{(\lambda)}]$ & $\begin{array}{ll}
     n^{1-2\lambda}& -\frac{1}{2}<\lambda<\frac{1}{2} \\ 
    \log{n}& \lambda=\frac{1}{2} \\
     \frac{3}{2\pi^{3/2}}\frac{\Gamma(\lambda-\frac{1}{2})}{\Gamma(\lambda)} & \lambda>\frac{1}{2}
\end{array}$& $\frac{\Gamma(2n+\frac{1}{2})}{\sqrt{2}\pi (n!)^{2}}\lambda^{1/2}$  \\ \hline
$  \mathcal{C}_{FS}[\hat{C}_n^{(\lambda)}]$ & $\begin{array}{ll}
      \frac{2\pi}{e^{3}} n^{3}& \lambda=\frac{1}{2} \\
     \frac{2^{1-4\lambda}(2\lambda-1)\pi} {e^{3}(\lambda^{2}-\lambda-3/4)} n^{3}& \lambda>\frac{3}{2} 
      \end{array}$& $\frac{2^{4n}(2n+1)}{(n!)^{4}\pi e}\lambda^{4n+1}$ \\ \hline
$\mathcal{C}_{LMC}[\hat{C}_n^{(\lambda)}]$ & $\begin{array}{ll}
      \frac{\pi 2^{1-2\lambda}}{e}n^{1-2\lambda}& -\frac{1}{2}<\lambda<\frac{1}{2} \\ 
     \frac{\pi}{e}\log{n}& \lambda=\frac{1}{2} \\ 
     \frac{2^{-2\lambda}}{e}\frac{3}{\sqrt{\pi}}\frac{\Gamma(\lambda-\frac{1}{2})}{\Gamma(\lambda)} & \lambda>\frac{1}{2} 
%      \\ & \\
%      \infty, & {\rm otherwise}.
      \end{array}$ & $\frac{2^{\frac{n-1}{2}}\Gamma(2n+\frac{1}{2})}{\pi (n!)^{5/2}}\lambda^{\frac{n+1}{2}}$ \\ \hline
\end{tabular}
\caption{First order asymptotics for the entropy-like ($F, \mathcal{L}_{S}, W_2$) and complexity-like ($\mathcal{C}_{FS},\mathcal{C}_{LMC}$) measures of the orthonormal Gegenbauer polynomials $\hat{C}_n^{(\lambda)}(x),$ $\lambda> -\frac{1}{2},\,\lambda\ne 0$,  when $n\to \infty$ and $\lambda\to \infty$. }
\label{Table_G}
\end{table}

\section{Fisher-Shannon complexity of the Laguerre polynomials}
\label{FisherShannonLag}

In this section we obtain simple analytical expressions for
 the Fisher-Shannon complexity of the orthonormal Laguerre polynomials $\hat{L}_n^{(\alpha)}(x)$ in the two following extreme situations: ($\alpha\rightarrow \infty; \mbox{fixed}\, n$) and ($n\rightarrow \infty; \mbox{fixed}\, \alpha$).
This quantity is defined (\ref{fishershannon}) as
\begin{equation}
     \label{fishershannonLag}
      \mathcal{C}_{FS}[\hat{L}_n^{(\alpha)}]=F[\hat{L}_n^{(\alpha)}] \times \frac{1}{2 \pi e} e^{2 S[\hat{L}_n^{(\alpha)}]}=\frac{1}{2 \pi e} F[\hat{L}_n^{(\alpha)}] \times \left(\mathcal{L}_{S}[\hat{L}_n^{(\alpha)}]\right)^2,
  \end{equation}
where the Fisher information has been shown to have the values \cite{Sanchez2005}
\begin{align}
       \label{fisher_l}
       F[L_n^{(\alpha)}]=
       \left\{
       \begin{array}{ll}
          4n+1, & \alpha=0, \\
          \frac{(2n+1)\alpha+1}{\alpha^2-1}, & \alpha>1, 
%          \\
%          \infty , & \alpha \in [-1,+1],\alpha \neq 0,
       \end{array}
       \right.
\end{align}
(being infinite otherwise, i.e., when $\alpha \in [-1,+1],\alpha \neq 0$), and the Shannon entropy power or Shannon spreading length $\mathcal{L}_{S}[\hat{L}_n^{(\alpha)}] = e^{S[\hat{L}_n^{(\alpha)}]}$ whose explicit expression is unknown despite multiple efforts (see e.g. the reviews \cite{Dehesa2001,Dehesa2021}). According to (\ref{shannon2}), the Shannon-like entropic functional $S[\hat{L}_n^{(\alpha)}]$ is given by
\begin{align}\label{SLI}
S\left[\hat{L}_n^{(\alpha)}\right]&=&-\int_{0}^{\infty}
\left[\hat{L}_n^{(\alpha)}(x)\right]^2h^{L}_{\alpha}(x) \log \left\{
\left[\hat{L}_n^{(\alpha)}(x)\right]^2 h^{L}_{\alpha}(x)\right\}dx
=E\left[\hat{L}_n^{(\alpha)}\right]+I\left[\hat{L}_n^{(\alpha)}\right],
\end{align}
with the integral functional \cite{Sanchez2011,Sanchez2000}
\begin{align}
	I\left[\hat{L}_n^{(\alpha)}\right] &=& -\int_{0}^{\infty}
\left[\hat{L}_n^{(\alpha)}(x)\right]^2 h^{L}_{\alpha}(x) \log h^{L}_{\alpha}(x)dx
 = 2n + \alpha+1-\alpha\,\psi(\alpha+n+1)
 \label{IntegralLag0}
\end{align}
(where $\psi(x)= \frac{\Gamma^{'}(x)}{\Gamma(x)}$ is the digamma function) and the Shannon entropy of $\hat{L}_n^{(\alpha)}(x)$ defined by
\begin{align}\label{eq_entropic_functional_laguerre}
E\left[\hat{L}_n^{(\alpha)}\right]&=-\int_{0}^{\infty}
\left[\hat{L}_n^{(\alpha)}(x)\right]^2 h^{L}_{\alpha}(x) \log
\left[\hat{L}_n^{(\alpha)}(x)\right]^2dx	.
\end{align}
The only existing approach to calculate this quantity requires the logarithmic potential of the Laguerre polynomials evaluated at their zeros, which is not analytically handy \cite{Dehesa2001}.\\

 Thus, the explicit expression of the Fisher-Shannon complexity of the Laguerre polynomials for generic values $(n,\alpha)$ is yet to be known. However, as shown below in this section and tabulated in Table \ref{Table_L}, there are two extremal situations where the value of this quantity can be expressed in a simple and transparent way: ($\alpha\rightarrow \infty; \mbox{fixed}\, n$) and ($n\rightarrow \infty; \mbox{fixed}\, \alpha$).
 
 \subsection{Asymptotics $\alpha\rightarrow \infty$}
To obtain the asymptotics ($\alpha\rightarrow \infty; \mbox{fixed}\, n$) of the Fisher-Shannon complexity $\mathcal{C}_{FS}[\hat{L}_n^{(\alpha)}]$, given by (\ref{fishershannonLag}), we first take into account from (\ref{fisher_l}) that $F[L_n^{(\alpha)}] \sim \frac{2n+1}{\alpha}$ and then, we determine the asymptotics the Shanon-like integral functional (\ref{SLI}) of the orthonormal Laguerre polynomials $\hat{L}_n^{(\alpha)}(x)$. 
To find the asymptotical ($\alpha\rightarrow \infty; \mbox{fixed}\, n$) value of $E\left[\hat{L}_n^{(\alpha)}\right]$ we express, following (\ref{E_pn}),  this quantity in terms of the corresponding one $E\left[L_n^{(\alpha)}\right]$ for the orthogonal polynomials 
\begin{equation}
	E\left[\hat{L}_n^{(\alpha)}\right] = \frac{1}{\kappa^{L}_{n,\alpha}} 	E[L_n^{(\alpha)}] + \log \kappa^{L}_{n,\alpha},
\end{equation}
%\textcolor{magenta}{From here below, there is an inconsistency in the notation, we have to choose $\log$ or $\ln$. }
and then we use the following asymptotical value  for the Shannon entropy of orthogonal Laguerre polynomials \cite{Belega2017,Temme2017}
\begin{equation}\label{ELaguerre}
	E[L_n^{(\alpha)}] \sim -\frac{\sqrt{2\pi}}{(n-1)!} \left(\frac{\alpha}{e}\right)^{\alpha}\,\alpha^{n+1/2}\,\log \alpha, \qquad \alpha\rightarrow \infty,
\end{equation}
and for the normalization constant $\kappa^{L}_{n,\alpha}$ given in the Table \ref{Table_HOPS}, which fulfills (keep in mind that $\Gamma(z) \sim e^{-z}\,z^z\,\left(\frac{2\pi}{z} \right)^{1/2}$, see Eq. 5.11.3 of \cite{Olver2010})
\begin{equation}
	\kappa^{L}_{n,\alpha} \sim  \kappa^{L}_{n,\infty} \equiv \frac{\sqrt{2\pi}}{n!} \left(\frac{\alpha}{e}\right)^{\alpha}\,\alpha^{n+1/2}, \qquad \alpha\rightarrow \infty,
	\label{kappa_constant_L}
\end{equation}
to finally obtain that 
\begin{equation}\label{ELI}
	E\left[\hat{L}_n^{(\alpha)}\right] =  \log\, \left(  \frac{\kappa^{L}_{n,\infty} }{\alpha^n }\right)=\log\, \left(\frac{\sqrt{2\pi\alpha}}{n!} \left(\frac{\alpha}{e}\right)^{\alpha}\right)+\mathcal{O}(\alpha^{-1}), \qquad \alpha\rightarrow \infty.
\end{equation}
%\textcolor{magenta}{I get that the $\alpha^{n}$ term goes in the denominator of Eq.23. This comes from the negative sign of eq.22}
The corresponding asymptotics for the functional $I\left[\hat{L}_n^{(\alpha)}\right]$ is given, according to (\ref{IntegralLag0}) and taking into account that $\psi(z) \sim \log z - \frac{1}{2z}$ for $z \rightarrow \infty$ (see Eq. 5.11.2 of \cite{Olver2010}) and $\alpha\log(\alpha+n+1) = \,\alpha\log(\alpha)+n+1+\mathcal{O}(\alpha^{-1})$, as 
\begin{equation}
	I\left[\hat{L}_n^{(\alpha)}\right] =  \log \left(\frac{e}{\alpha}\right)^{\alpha}+n+\frac{1}{2}+\mathcal{O}(\alpha^{-1}), \qquad \alpha\rightarrow \infty.
	\label{I_Shannon_L}
\end{equation}

%\textcolor{magenta}{Note: This step Eq.(\ref{I_Shannon_L}) is non-trivial, one has to take into account that}

%\begin{align}
%\textcolor{magenta}{\alpha\log(n+\alpha+1)\sim\,\alpha\log(\alpha)+n+1+\mathcal{O}(\alpha^{-1/2})\,; \qquad \alpha\rightarrow \infty}
%\end{align}
%\textcolor{magenta}{where it has been used that $(\alpha+b)^{\alpha}\sim\alpha^{\alpha}+\alpha^{\alpha}b$ with $b=n+1$ (Newton Binomia) and $log(1+n)\sim n$ (checked numerically. The justification of adding $\mathcal{O}(1)$ terms in Eq.(\ref{I_Shannon_L}) is based on the fact that the leading term in the final expression for $C_{FS}[\hat{L}_n^{(\alpha)}]$ is exactly $\mathcal{O}(1)$.}\\
%NAHUAL, OJO, $log(1+n)\sim n$ NOT POSSIBLE, porque n es finito, fijo. Creo que $\log(\alpha+b) \sim \log \alpha+ \frac{b}{\alpha} + \mathcal{O}(\alpha^{-2})$, de forma que con $b=n+1$ se tiene
%\textcolor{magenta}{ Toda la razon. Comprobado analitica y numericamente.}
%y por tanto la Eq. (25) es 
%\begin{equation}
%	I\left[\hat{L}_n^{(\alpha)}\right] \sim  \log \left(\frac{e}{\alpha}\right)^{\alpha}\textcolor{magenta}{+n+1+\mathcal{O}(\alpha^{-1})} ; \qquad \alpha\rightarrow \infty
%	\label{I_Shannon_L}
%\end{equation}
%ESTO afectara a lo que sigue. \textcolor{magenta}{Jesus, sigo creyendo que el resultado de Eq. (25) es el correcto y no Eq. (27). El termino de 1/2 proviene de el factor 1/(2z) de la aproximacion de la funcion digamma (el 1 de la Eq. (26) se cancela con el 1 de la Eq. (19)). Comprobado numericamente tambien.}

Then, from Eqs. (\ref{SLI}), (\ref{ELI}) and (\ref{I_Shannon_L}), we have that the asymptotics ($\alpha\rightarrow \infty; \mbox{fixed}\, n$) of the Shannon entropy power of the Laguerre polynomials $\mathcal{L}_{S}[\hat{L}_n^{(\alpha)}]$ is given by
\begin{equation}\label{shaentroasy}
	\mathcal{L}_{S}[\hat{L}_n^{(\alpha)}] = e^{S[\hat{L}_n^{(\alpha)}]} = \,\left(
 \frac{e}{\alpha} \right)^\alpha\,  \frac{\kappa^{L}_{n,\infty}}{\alpha^{n}} e^{n+1/2} = \frac{\sqrt{2\pi\alpha}}{n!}e^{n+1/2}+\mathcal{O}(\alpha^{-1/2})\,, \qquad \alpha\rightarrow \infty.
 \end{equation}
% \textcolor{magenta}{From my derivation (in agreement with the new Eq.24), the Shannon entropic power encreases as $\sqrt{\alpha}$ and not as $\alpha^{2n+1/2}$.}
 Finally, according to Eqs. (\ref{fishershannonLag}), (\ref{fisher_l}) and (\ref{shaentroasy}), we have that the Fisher-Shannon complexity of the Laguerre polynomials behaves as
 \begin{equation}\label{CFSLAG}
 	\mathcal{C}_{FS}[\hat{L}_n^{(\alpha)}] = \frac{2n+1}{(n!)^2}e^{2n}+\mathcal{O}(\alpha^{-1}) , \qquad \alpha\rightarrow \infty.
 \end{equation}
We observe that the first dominant term does not depend on the Laguerre parameter, indicating uniformity (perfect disorder) for the Rakhmanov probability. This happens because the two entropic components of structure-order (Fisher information) and disorder (Shannon entropic power) qualitatively cancel when $\alpha\rightarrow \infty$.  Eventually, we can go further away by obtaining the second asymptotical term. This requires to improve the asymptotical behavior of the Shannon entropy (\ref{ELaguerre}) of the Laguerre polynomials, what it is a feasible task following the lines of \cite{Belega2017,Temme2017}. The latter may be an interesting task for the future, not only in the theory of orthogonal polynomials but also for its physical consequences. Indeed, the determination of the complexity of the charge distribution of the quasi-classical (i.e., high-dimensional) states of quantum systems with a spherically-symmetric potential boils down to the  mathematical computation of the complexity of the Laguerre polynomials. This is because the radial eigenfunction of the quasi-classical states are controlled by Laguerre polynomials with a parameter $\alpha$ which linearly depends on the space dimensionality of the system \cite{Puertas2017,Puertas2017b,Dehesa2019,Dehesa2020}.

 \subsection{Asymptotics $n\rightarrow \infty$}

 In addition, for completeness, let us briefly show that the asymptotics ($n\rightarrow \infty; \mbox{fixed}\, \alpha$) of the Laguerre polynomials is known \cite{Sanchez2011} to behave as
 \begin{align}\label{CFSLag}
      \mathcal{C}_{FS}\left[\hat{L}_n^{(\alpha)}\right]\sim
      \left\{
      \begin{array}{ll}
      \left(\frac{8 \pi}{e^3}\right) n^{3},& \alpha=0, \\ & \\
      \frac{\alpha}{\alpha^2-1} \left(\frac{4\pi}{e^3}\right) n^{3}, & \alpha>1, 
%      \\ & \\
%      \infty, & {\rm otherwise}.
      \end{array}
      \right.
\end{align}
Basically, this is because the Fisher information is given by (\ref{fisher_l}) and the Shannon-like entropic functional $S[\hat{L}_n^{(\alpha)}]$ has the non-trivial value \cite{Dehesa2001}
\[
S[\hat{L}_n^{(\alpha)}]=(\alpha+1)\log n-\alpha \psi(\alpha+n+1)-1+\log(2\pi)+\mathcal{O}(1),\qquad n \rightarrow \infty,
\]
so that the Shannon entropic power $\mathcal{L}_{S}[L_n^{(\alpha)}]$ fulfills that
\begin{equation}
      \label{shannon_length}
      \mathcal{L}_{S}[\hat{L}_n^{(\alpha)}] \sim \frac{2 \pi}{e}\,n,\qquad n\rightarrow \infty.
\end{equation}
Note from (\ref{CFSLag}) that  the Fisher-Shannon complexity of the Laguerre polynomials follows a growth scaling law $n^3$ when $n\rightarrow \infty$, because the Fisher and Shannon components combine constructively since they behave as $n$ and $n^2$ for ($n\rightarrow \infty$; fixed $\alpha$), respectively. Interestingly, this is specially useful to explain the charge complexity of highly-excited (Rydberg) states of the multidimensional Coulomb and oscillator-type systems \cite{Aptekarev2016,Dehesa2017,Dehesa2020,Dehesa2021,Aptekarev2021}. Basically, this is because the radial eigenfunctions of these multidimensional quantum systems are controlled by Laguerre polynomials \cite{Yanez1994,Dehesa2001}.

\section{LMC complexity of the Laguerre polynomials}
\label{LMCLag}
In this section we obtain simple and compact expressions in two extreme situations, ($\alpha\rightarrow \infty; \mbox{fixed}\, n$) and ($n\rightarrow \infty; \mbox{fixed}\, \alpha$), for
 the LMC complexity of the orthonormal Laguerre polynomials $\hat{L}_n^{(\alpha)}(x)$. This quantity, according to (\ref{lmc}), is given by
   \begin{equation}\label{lmcdef}
      \mathcal{C}_{LMC}[\hat{L}_n^{(\alpha)}] = W_2[\hat{L}_n^{(\alpha)}] \times \mathcal{L}_{S}[\hat{L}_n^{(\alpha)}].
   \end{equation}
The explicit expressions of this quantity at generic values of $n$ and $\alpha$ is not yet known, although there are highbrow, non-handy analytical expressions for the second-order entropic moment $W_2[\hat{L}_n^{(\alpha)}]$ and the Shannon entropic power $\mathcal{L}_{S}[\hat{L}_n^{(\alpha)}]$ which allow one to calculate them in an algorithmically symbolic manner. In fact, the computation of $W_2[\hat{L}_n^{(\alpha)}]$ requires \cite{Sanchez2011,Dehesa2015} the evaluation of the four-variate Lauricella function $F_A^{(4)}\left(\frac12,\frac12,\frac12,\frac12\right)$ or the computation of the multivariate Bessel polynomials of combinatorics evaluated at the expansion coefficients of the Laguerre polynomials; and the computation of the Shannon entropic power $\mathcal{L}_{S}[\hat{L}_n^{(\alpha)}]$ requires \cite{Dehesa2001} the evaluation of the logarithmic potential of the Laguerre polynomials at their zeros.\\

\subsection{Asymptotics $\alpha\rightarrow \infty$}

To obtain the asymptotics ($\alpha\rightarrow \infty; \mbox{fixed}\, n$) of the LMC complexity $\mathcal{C}_{LMC}[\hat{L}_n^{(\alpha)}]$ we begin with the asymptotical expression (\ref{shaentroasy}) of $\mathcal{L}_{S}[\hat{L}_n^{(\alpha)}]$ already shown in the previous section. Let us now tackle the asymptotics for the the second-order entropic moment $W_2[\hat{L}_n^{(\alpha)}]$ given by
\begin{equation}
	W_2[\hat{L}_n^{(\alpha)}] = \int_{0}^{\infty}
\left(\left[\hat{L}_n^{(\alpha)}(x)\right]^2 h^{L}_{\alpha}(x)\right)^2\,dx = \int_{0}^{\infty} \,x^{2\alpha}\,e^{-2\alpha}\,\left[\hat{L}_n^{(\alpha)}(x)\right]^4\,dx
\end{equation}
Now, we use the recent methodology of Temme et al \cite{Temme2017}. Let $\alpha, \lambda, \kappa$, and $\mu$ be positive real numbers; then, the following R\'{e}nyi-like functional for orthogonal Laguerre polynomials fulfills the asymptotics
\begin{equation}
\label{eq:Lag19}
\int_{0}^{\infty}x^{\mu-1}e^{-\lambda x} \left|L_{m}^{(\alpha)}(x)\right|^{\kappa}\,dx \sim   \frac{\alpha^{\kappa m}\Gamma(\mu)}{\lambda^\mu (m!)^\kappa}, \qquad \alpha \to\infty. 
\end{equation}
Then, with the values $\mu= 2\alpha+1,\, \lambda=2$ and $\kappa = 4$, this general asymptotical formula provides the required asymptotics for $W_2[\hat{L}_n^{(\alpha)}]$:
\begin{equation}\label{omega2L}
	W_2[\hat{L}_n^{(\alpha)}] \sim \frac{1}{(k_{n,\infty}^{L})^{2}} \frac{\alpha^{4n}\,\Gamma(2\alpha+1)}{2^{2\alpha+1}\,(n!)^4}, \qquad \alpha \to\infty.
\end{equation}
Using now the (previously given) asymptotical estimate for the gamma function together with Eq. (\ref{kappa_constant_L}), one finds 
\begin{equation}\label{omega2L}
	W_2[\hat{L}_n^{(\alpha)}] =\alpha^{2n}\left(\frac{1}{2\,(n!)^{2}\sqrt{\pi \alpha}}+\mathcal{O}(\alpha^{-3/2})\right), \qquad \alpha \to\infty.
\end{equation}
Finally, the combination of Eqs. (\ref{lmcdef}), (\ref{shaentroasy}) and (\ref{omega2L}) lead us to the following asymptotical values of the LMC complexity of the Laguerre polynomials:
\begin{equation}\label{CLMCLag}
	 \mathcal{C}_{LMC}[\hat{L}_n^{(\alpha)}] = \alpha^{2n}\left(\frac{e^{n+1/2}}{2^{1/2}(n!)^{3}}+\mathcal{O}(\alpha^{-\frac{1}{2}})\right), \qquad \alpha \to\infty.
\end{equation}
Note that this quantity behaves as $\alpha^{2n}$ when $\alpha\rightarrow \infty$ because its two order (entropic moment $W_2[\hat{L}_n^{(\alpha)}]$) and disorder (Shannon spreading length $\mathcal{L}_{S}[\hat{L}_n^{(\alpha)}]$) components contribute constructively as $(\alpha^{2n-1/2},\alpha^{1/2})$ at first asymptotical order.
In fact, this expression can be improved by using higher terms in the  asymptotical expression (\ref{eq:Lag19}) following the method of Temme et al. \cite{Temme2017}. This is relevant \textit{per se} and because this quantity allows us to determine the corresponding statistical complexity of the high-dimensional states of both multidimensional hydrogenic and oscillator systems. The latter is because the Laguerre polynomials control the radial eigenfunctions of the high-dimensional states of these quantum systems \cite{Puertas2017,Puertas2017b,Dehesa2019,Dehesa2020} as previously mentioned.

\subsection{Asymptotics $n\rightarrow \infty$}

Let us now tackle the asymptotics $(n\rightarrow \infty; \mbox{fixed}\, \alpha)$ of the LMC complexity of the Laguerre polynomials. Then, we take into account the asymptotical value (\ref{shannon_length}) for the Shannon entropic power $\mathcal{L}_{S}[L_n^{(\alpha)}]$, and to find the corresponding asymptotics of $W_2[\hat{L}_n^{(\alpha)}]$ we use the recent asymptotics for the generalized weighted $\mathfrak{L}_{q}$-norms of Aptekarev et al. \cite{Aptekarev2016} which, in particular, gives
\begin{equation}\label{omega2inf}
	W_2[\hat{L}_n^{(\alpha)}] \sim \frac{\log n+\mathcal{O}(1)}{\pi^{2}n},\qquad n\rightarrow \infty.
\end{equation}
Finally, according to (\ref{shannon_length}), (\ref{lmcdef}) and (\ref{omega2inf}), we obtain the following asymptotics for the LMC complexity of orthonormal Laguerre polynomials
\begin{equation}
	\mathcal{C}_{LMC}[\hat{L}_n^{(\alpha)}] = \frac{2}{\pi\,e}\,\log n+\mathcal{O}(n^{-1}),\qquad n\rightarrow \infty.
\end{equation}
Thus, the LMC complexity of the Laguerre polynomials follows a logarithmic growth scaling law at large degree $n$; basically, because its two entropic components $(W_2,\mathcal{L}_{S})$  behave as $(\frac{\log n}{n}, n)$, respectively. This mathematical result allows us to compute the corresponding radial charge complexity for the Rydberg quantum states of the hydrogenic and harmonic systems, because the radial eigenfunctions of such states are controlled by the Laguerre polynomials \cite{Yanez1994,Dehesa2001}. \\

\section{Fisher-Shannon complexity of the Gegenbauer polynomials}
\label{FisherShannonGeg}
In this section we obtain the Fisher-Shannon complexity (\ref{fishershannon}) of the orthonormal Gegenbauer polynomials $\hat{C}_n^{(\lambda)}(x), \lambda> -\frac{1}{2}$ when ($\alpha\rightarrow \infty; \mbox{fixed}\, n$) and for ($n\rightarrow \infty; \mbox{fixed}\, \alpha$). This quantity is defined as
\begin{equation}
     \label{fishershannonGeg}
      \mathcal{C}_{FS}[\hat{C}_n^{(\lambda)}]=F[\hat{C}_n^{(\lambda)}] \times \frac{1}{2 \pi e} e^{2 S[\hat{C}_n^{(\lambda)}]}=\frac{1}{2 \pi e} F[\hat{C}_n^{(\lambda)}] \times \left(\mathcal{L}_{S}[\hat{C}_n^{(\lambda)}] \right)^2,
  \end{equation}
  The explicit expression of the Fisher-Shannon complexity of the Gegenbauer polynomials for generic values $(n,\,\lambda)$ is unknown up until now, basically because the Shannon entropy is also not known despite many efforts \cite{Buyarov2000,Vicente2007}. However, there are two extremal situations where the value of this quantity can be analytically expressed when ($\lambda\rightarrow \infty; \mbox{fixed}\, n$) and when ($n\rightarrow \infty; \mbox{fixed}\, \lambda$). The goal of this section is to obtain both the parameter and degree asymptotics in a compact way.\\
  
  The Fisher information of the Gegenbauer polynomials $F[\hat{C}_n^{(\lambda)}]$ can be obtained from the corresponding quantity $F\left[\hat{P}_n^{(\alpha,\beta)}\right]$ of the Jacobi polynomials $\hat{P}_n^{(\alpha,\beta)}(x)$, given \cite{Sanchez2005,Guerrero2010} by
\begin{align}\label{fisherjacobi}
F\left[\hat{P}_n^{(\alpha,\beta)}\right]=
\left\{
\begin{array}{ll}
2n(n+1)(2n+1), & \alpha,\beta=0, \\
\frac{2n+\beta+1}{4}\left[\frac{n^2}{\beta+1}+n+(4n+1)(n+\beta+1)+\frac{(n+1)^2}{\beta-1}\right],
& \alpha=0, \beta>1, \\
\frac{2n+\alpha+\beta+1}{4(n+\alpha+\beta-1)}\left[n(n+\alpha+\beta-1)
\left(\frac{n+\alpha}{\beta+1}+2+\frac{n+\beta}{\alpha+1}\right)\right.& \\
+\left.(n+1)(n+\alpha+\beta)\left(\frac{n+\alpha}{\beta-1}+2+\frac{n+\beta}{\alpha-1}\right)\right],
&\alpha, \beta>1, \\
\infty, & {\rm otherwise.}
\end{array}
\right.
\end{align}
From this expression and taking into account the relation of the orthogonal/orthonormal Gegenbauer polynomials and the Jacobi polynomials given as
  \begin{equation}
  	\hat{C}_n^{(\lambda)}(x) = (\kappa_{n,\lambda}^{G})^{-\frac{1}{2}}\,C_n^{(\lambda)}(x),
  \end{equation} 
  \begin{equation}
  	C_n^{(\lambda)}(x) =c_{n,{\lambda}}P_{n}^{(\lambda-\frac{1}{2},\lambda-\frac{1}{2})}(x)\equiv \frac{\Gamma (\lambda+\frac{1}{2})}{\Gamma(2\lambda)}\,\frac{\Gamma(n+2\lambda)}{\Gamma(n+\lambda+\frac{1}{2})}\,P_{n}^{(\lambda-\frac{1}{2},\lambda-\frac{1}{2})}(x),
  \end{equation}
%  \textcolor{magenta}{I find convenient to introduce the constant $c_{n,{\lambda}}$ that relates the Gegenbauer with the Jacobi in order to see in a more natural way the Fisher information, see Eq $F\left[\hat{C}_n^{(\lambda)}\right]$}
together with the exact identity $c_{n,{\lambda}}\left(\kappa^{J}_{n,\lambda-1/2,\lambda-1/2}/\kappa^{G}_{n,\lambda}\right)^{1/2}=1$, we have that the values of the Fisher information of the Gegenbauer polynomials are found to be
\begin{align}\label{FG}
F\left[\hat{C}_n^{(\lambda)}\right]=F\left[\hat{P}_n^{(\lambda-1/2,\lambda-1/2)}\right]=
\left\{
\begin{array}{ll}
2n(n+1)(2n+1), & \lambda = \frac{1}{2}, \\
%(n+\lambda)\left[\frac{n(n+2\lambda+1)}{\lambda+1/2}
%+ \frac{(n+1)(n+2\lambda-1)}{\lambda-3/2}\right], THIS EXPRESSION IS WRONG
\frac{2(n+\lambda)(2\lambda-1)\left(1+2\lambda+2n(n+2\lambda)\right)}{(2\lambda-3)(1+2\lambda)}
&\lambda > \frac{3}{2}, \\
\infty, & {\rm otherwise.}
\end{array}
\right.
\end{align}
%\textcolor{magenta}{The term in blue is exactly =1 (as it should be).Checked analitically.}
  
  In addition, the Shannon entropic power $\mathcal{L}_{S}[\hat{C}_n^{(\lambda)}] = e^{S[\hat{C}_n^{(\lambda)}]}$   where the Shanon-like integral functional of the orthonormal Gegenbauer polynomials $\hat{C}_n^{(\lambda)}(x)$, according to (\ref{shannon2}), is given by
\begin{align}\label{shannonGeg}
S\left[\hat{C}_n^{(\lambda)}\right]&=&-\int_{-1}^{+1}
\left[\hat{C}_n^{(\lambda)}(x)\right]^2h^{G}_{\lambda}(x) \log\left\{
\left[\hat{C}_n^{(\lambda)}(x)\right]^2 h^{G}_{\lambda}(x)\right\}dx
=E\left[\hat{C}_n^{(\lambda)}\right]+I\left[\hat{C}_n^{(\lambda)}\right],
\end{align}
with the integral functional \cite{Sanchez2000}
\begin{align} \label{IntegralGeg0}
	I\left[\hat{C}_n^{(\lambda)}\right] &= -\int_{-1}^{+1}
\left[\hat{C}_n^{(\lambda)}(x)\right]^2 h^{G}_{\lambda}(x) \log h^{G}_{\lambda}(x)\nonumber\\
 &= \frac{(2\lambda-1)\pi n!\Gamma(n+2\lambda)}{2^{2(n+\lambda)-1}(n+\lambda)[\Gamma(n+\lambda)]^{2}}\left(\frac{1}{2(n+\lambda)}+\log(2)+\psi(n+\lambda)-\psi(n+2\lambda)\right)
 \end{align}
 and the Shannon entropy of $\hat{C}_n^{(\lambda)}(x)$ is defined by
\begin{align}\label{eq_entropic_functional_laguerre}
E\left[\hat{C}_n^{(\lambda)}\right]&=-\int_{-1}^{+1}
\left[\hat{C}_n^{(\lambda)}(x)\right]^2 h^{G}_{\lambda}(x) \log
\left[\hat{C}_n^{(\lambda)}(x)\right]^2dx .
\end{align}
The analytical determination of the latter quantity is a formidable task \cite{Buyarov2000,Vicente2007}. Indeed, it has have been  calculated for integer values of the polynomial's parameter and in a somewhat highbrow manner \textit{only}. However, we find below that they can be expressed in a simple and compact way in the two following extremal situations; the main results have been collected in Table  \ref{Table_G}. \\

%Nota: To obtain both parameter and degree asymptotic expressions pay attention to \cite{Buyarov2000,Vicente2007,Dehesa2001}. Ver sobre todo la seccion  5 del paper review Dehesa2001 \cite{Dehesa2001} donde se da la relacion entre las entropias de los polinomios ortonormales y los ortogonales de Gegenbauer con la entropia de los polinomios de Gegenbauer  $E({\hat{G}}^{\alpha}_{m})$ de cuya asintotica ($n\rightarrow \infty; \text{fixed}\, \alpha$) se da tambien el termino dominante cuando $\alpha$ es entero; tambien se da el segundo termino pero en una forma poco analytical y nada handy. Ver en este sentido tambien el paper de Buyarov2000 \cite{Buyarov2000}. Digamos tambien que en el paper de Vicente2007 \cite{Vicente2007} se da una expresion analitica para cualquier $\alpha$ entero peo lamentablemente tampoco es analiticamente muy manejable.

  \subsection{Asymptotics $n\rightarrow \infty$}
  
  %  From (\ref{fisherjacobi}) the Fisher information of the orthonormal Jacobi polynomials $F[\hat{P}_n^{(\alpha,\beta)}]$ has the following asymptotical ($n\rightarrow \infty; \text{fixed}\, \alpha$) values
%\begin{align*}
%F\left[\hat{P}_n^{(\alpha,\beta)}\right]=
%\left\{
%\begin{array}{ll}
%4\,n^3 + \mathcal{O}(n^2), & \alpha,\beta=0, \\
%\frac{2\beta^2+\beta-2}{\beta^2-1}\,n^3 + \mathcal{O}(n^2),
%& \alpha=0, \beta>1, \\
%\frac{(\alpha \beta - 1)(\alpha+\beta)}{(\alpha^2-1)(\beta^2-1)}\,n^3 + \mathcal{O}(n^2) ,
%&\alpha, \beta>1, \\
%\infty, & {\rm otherwise,}
%\end{array}
%\right.
%\end{align*}
%\textcolor{magenta}{The above eq has been correctly checked numerically. }
From expression (\ref{FG}) we can obtain the following asymptotics ($n\rightarrow \infty; \mbox{fixed}\, \lambda$) behavior for the Fisher information of the Gegenbauer polynomials $F[\hat{C}_n^{(\lambda)}]$ :
\begin{align}\label{ANFSG}
F\left[\hat{C}_n^{(\lambda)}\right]=
\left\{
\begin{array}{ll}
4\,n^3 + \mathcal{O}(n^2), & \lambda = \frac{1}{2}, \\
\frac{2\lambda - 1}{\lambda^2-\lambda-\frac{3}{4}}\,n^3 + \mathcal{O}(n^2) ,
&\lambda > \frac{3}{2}, \\
\infty, & {\rm otherwise,}
\end{array}
\right.
\end{align}
The asymptotics of $\mathcal{L}_{S}[\hat{C}_n^{(\lambda)}]$  requires to find the asymptotics of the Shannon entropy-like functional $S\left[\hat{C}_n^{(\lambda)}\right]$ which, according to  (\ref{shannonGeg}), involves the asymptotics ($n\rightarrow \infty; \mbox{fixed}\, \lambda$) of the Shannon entropy $E\left[\hat{C}_n^{(\lambda)}\right]$ and the integral functional $I\left[\hat{C}_n^{(\lambda)}\right]$ given by (\ref{IntegralGeg0}). 
% At times it is convenient the use of the polynomials 
% \begin{align*}
%	G_{k}^{\lambda}(x) &=& \left( \frac{k! (k+\lambda) \Gamma(2 \lambda)}{\lambda \Gamma(k+2 \lambda)} \right)^{\frac{1}{2}} C_{k}^{\lambda}(x) \\ 
%& = &\gamma_{k}^{\lambda}  x^k+ \mbox{lower degree terms}
%\end{align*}
%which are orthogonal with respect to the positive unit weight on $[-1,+1]$,
%\begin{equation*}
%\omega_\lambda = \frac{\Gamma(\lambda+1)}{\sqrt{\pi} \Gamma(\lambda+1/2)} (1-x^2)^{\lambda-\frac{1}{2}}
%\end{equation*}
% Then, \underline{the Shannon entropy of these Gegenbauer polynomials} is given by
%\begin{equation*}
%E \left[G_{k}^{\lambda} \right] \equiv - \int_{-1}^{+1} \left[G_{k}^{\lambda}(x) \right] \log \left[G_{k}^{\lambda}(x) \right] \omega_{\lambda}(x) dx
%\end{equation*}
The Shannon entropy of $\hat{C}_n^{(\lambda)}(x)$ has the following degree asymptotical behavior
\cite{Aptekarev1994,Aptekarev1996,Dehesa2001}:
\begin{align}
E({\hat{C}}^{(\lambda)}_{n})& \equiv -\int^{+1}_{-1} h_\lambda (x) \left[ \hat{C}^{(\lambda)}_{n}(x) \right]^2 \log\left[ \hat{C}^{(\lambda)}_{n}(x) \right]^2 dx\nonumber\\\label{SEC}
&= \log \pi+(1-2 \lambda) \log 2-1+\mathcal{O}(n^{-1}),\qquad n\rightarrow \infty
\end{align}
for fixed $\lambda$ \cite{Aptekarev1994,Dehesa2001}. Moreover, from (\ref{IntegralGeg0}) and the previously given asymptotical expressions for the gamma and digamma functions, we find that the functional $I\left[\hat{C}_n^{(\lambda)}\right]$ behaves for fixed $\lambda$ as 
\begin{align}\label{IGegninf}
I\left[\hat{C}_n^{(\lambda)}\right] = 2^{1-2(n+\lambda)}(\pi(2\lambda-1)\log 2+\mathcal{O}(n^{-\frac{1}{2}})),\qquad n\rightarrow \infty
\end{align}

%The $2^{-2n}$ contribution in \ref{IGegninf} let us conclude that in the limit $n\rightarrow \infty$ we can approximate $S\left[\hat{C}_n^{(\lambda)}\right]\sim E\left[\hat{C}_n^{(\lambda)}\right]$.

 Therefore, from (\ref{shannonGeg}), (\ref{SEC}) and (\ref{IGegninf}) we find that the asymptotics for the Shanon-like functional of the Gegenbauer polynomials is
\begin{align}
S\left[\hat{C}_n^{(\lambda)}\right]\sim E\left[\hat{C}_n^{(\lambda)}\right]=\log \pi+(1-2 \lambda) \log 2-1+\mathcal{O}(n^{-1}),\qquad n\rightarrow \infty,
\end{align}
so that the Shannon entropy power has the behavior
\begin{equation}\label{ANLSG}
	\mathcal{L}_{S}[\hat{C}_n^{(\lambda)}] \sim \frac{\pi\, 2^{1-2\lambda}}{e}, \qquad n\rightarrow \infty
\end{equation}

Finally, taking into account (\ref{fishershannonGeg}), (\ref{ANFSG}) and (\ref{ANLSG}) we have that the Fisher-Shannon complexity for the orthonormal Gegenbauer polynomials has the expression
 \begin{align}\label{FSGEG}
      \mathcal{C}_{FS}\left[\hat{C}_n^{(\lambda)}\right]=
      \left\{
      \begin{array}{ll}
      \frac{2\pi}{e^{3}} n^{3}+\mathcal{O}(n^{2}),& \lambda=\frac{1}{2}, \\ & \\
     \frac{2^{1-4\lambda}(2\lambda-1)\pi} {e^{3}(\lambda^{2}-\lambda-3/4)} n^{3}+\mathcal{O}(n^{2}), & \lambda>\frac{3}{2}, 
%      \\ & \\
%      \infty, & {\rm otherwise}.
      \end{array}
      \right.
\end{align}
in the limit $n\rightarrow \infty$. Further terms can be obtained by improving the asymptotics (\ref{IGegninf}) of the Shannon entropy $E\left[\hat{C}_n^{(\lambda)}\right]$ as previously indicated \cite{Buyarov2000,Vicente2007}. Note that the Fisher-Shannon complexity of the Gegenbauer polynomials behaves dominantly according to the scaling law $n^3$ for large degrees $n$; so, similarly to the Laguerre case (see (\ref{CFSLag})) but for different reasons. Indeed, the entropic Fisher and Shannon components behave according to laws $(n^3, constant)$ and $(n, n^2)$ for the Gegenbauer and Laguerre cases, respectively. This indicates that when $n\rightarrow \infty$, the gradient content is much higher for the Gegenbauer polynomials  than for the Laguerre polynomials, while the disequilibrium (i.e., deviation from the uniform distribution) in the Gegenbauer case is much lower than in the Laguerre case for any fixed degree.\\

Finally, let us mention that expression (\ref{FSGEG})  allows one to compute the corresponding radial momentum complexity for the Rydberg quantum states of the hydrogenic and harmonic systems, because the radial eigenfunctions of such states are controlled by the Gegenbauer polynomials \cite{Yanez1994,Dehesa2001}.

\subsection{Asymptotics $\lambda\rightarrow \infty$}
\label{asympspreading}

Let us now determine the LMC complexity $\mathcal{C}_{FS}[\hat{C}_n^{(\lambda)}]$ in the limit $\lambda\rightarrow \infty$ with fixed degree $n$. For this purpose we first make use of Temme et al.'s ideas \cite{Temme2017} to derive the Shannon entropy of $\hat{C}_n^{(\lambda)}(x)$ from the corresponding asymptotics of the $\mathcal{N}_{p}$-norm  of the orthogonal Gegenbauer polynomials, defined as
\begin{align}\label{Np_norm}
\mathcal{N}_{p}\left[C_n^{(\lambda)}\right]=\int_{-1}^{1}(1-x^{2})^{\lambda-\frac{1}{2}}\left|C_n^{(\lambda)}\right|^{p}dx.
\end{align}

This quantity can be analytically estimated for $\lambda\to\infty$ by taking into account the known relation \cite{Olver2010}
\begin{equation}\label{limitGeg}
	\lim_{\lambda \rightarrow \infty} \frac{C_n^{(\lambda)}(x)}{C_n^{(\lambda)}(1)} = x^n,
\end{equation}
with
\begin{equation}
	C_n^{(\lambda)}(1) = \binom {n+2\lambda-1}{n} = \frac{(n+2\lambda-1)!}{n! \,(2\lambda-1)!}.
\end{equation}
Then, from (\ref{Np_norm}) and (\ref{limitGeg}) we have

\begin{align}\label{Np_norm_2}
\mathcal{N}_{p}\left[C_n^{(\lambda)}\right]\sim \left[C_n^{(\lambda)}(1)\right]^{p}\frac{\Gamma(\frac{1}{2}(1+np))\Gamma(\frac{1}{2}+n)}{\Gamma(1+\lambda+\frac{np}{2})}.
\end{align}
Now, according to Eqs. (\ref{E_p}) and (\ref{Np_norm_2}), one has that the Shannon entropy  of the orthogonal $C_n^{(\lambda)}(x)$ in the current limit is given as
\begin{eqnarray}
	E\left[C_n^{(\lambda)}\right] =2 \frac{d}{dp}\left[\mathcal{N}_{p}\left[C_n^{(\lambda)}\right]\right]_{p=2}\nonumber\\
	\sim 2\,\kappa_{n,\lambda}^{G} \left(\log \left[\frac{(n+2\lambda-1)!}{n! \,(2\lambda-1)!}\right] + \frac{n}{2}\, \psi(\frac{2n+1}{2})-\frac{n}{2}\,\psi(n+2\lambda+1)\right),
\end{eqnarray}

so that we can express the Shannon entropy  of  the orthonormal Gegenbauer polyomials as
\begin{gather}
	E\left[\hat{C}_n^{(\lambda)}\right] \sim 2\left(\log \left[\frac{(n+2\lambda-1)!}{n! \,(2\lambda-1)!}\right] + \frac{n}{2}\,\psi(\frac{2n+1}{2})-\frac{n}{2}\,\psi(n+2\lambda+1)\right)
	\nonumber\\
	\sim 2\log\left(\frac{\lambda^{n}2^{n}}{n!}\right).
	%+\mathcal{O}(\log(\lambda^{n-1}))
	\label{E_lambda_infinity}
\end{gather}
%Primero habra que obtener la asintotica ($\lambda\rightarrow \infty; \text{fixed}\, n$) de la Fisher.
%En cuanto al otro factor $\mathcal{L}_{S}[\hat{C}_n^{(\lambda)}]$, que involucra $S\left[\hat{C}_n^{(\lambda)}\right]$ y por tanto $E\left[\hat{C}_n^{(\lambda)}\right]$ y $I\left[\hat{C}_n^{(\lambda)}\right]$. La asintotica de $I\left[\hat{C}_n^{(\lambda)}\right]$ puede obtenerse de la Eq. (\ref{IntegralGeg0}). Y la asintotica de $E\left[\hat{C}_n^{(\lambda)}\right]$ creo que la podemos resolver con la Nota 2.pdf, y quizas tambien con la seccion 4 del paper Temme1 \cite{Temme2017} antes citado. En efecto, se obtiene 
%(hay que chequear y terminar esto. Dar algunos pasos intermedios mas). Mencionemos tambien por chequeo que para los polinomios ortogonales, se tiene que 
%\begin{equation}
%	E\left[C_n^{(\lambda)}\right] \sim \frac{1}{2}\,\kappa_{n,\lambda}^{G} \left(\log \left[\frac{(n+2\lambda-1)!}{n! \,(2\lambda-1!}\right] + \frac{n}{2}\, \psi(\frac{2n+1}{2})-\frac{n}{2}\,\psi(n+2\lambda+1)\right)
%\end{equation}
In addition, the integral functional $I\left[\hat{C}_n^{(\lambda)}\right]$ given by (\ref{IntegralGeg0}) behaves as
\begin{align}\label{ECn}
I\left[\hat{C}_n^{(\lambda)}\right]=\lambda^{-n}\left(\frac{(2 n+1) n!}{2^{1 + n}}\sqrt{\frac{\pi}{\lambda}}+\mathcal{O}(\lambda^{-3/2})\right),\qquad \lambda\rightarrow \infty.
\end{align}
Then, according to Eqs. (\ref{shannonGeg}), (\ref{E_lambda_infinity}) and (\ref{ECn}), we find the following asymptotics for the Shannon-like functional of the Gegenbauer polynomials
\begin{align}
\label{SC_lambda_infinity}
S\left[\hat{C}_n^{(\lambda)}\right]\sim E\left[\hat{C}_n^{(\lambda)}\right]\sim 2\log\left(\frac{\lambda^{n}2^{n}}{n!}\right),\qquad  \lambda\rightarrow \infty,
\end{align}
so that the Shannon entropy power of Gegenbauer polynomials behaves as 
\begin{equation}\label{LSCn}
	\mathcal{L}_{S}[\hat{C}_n^{(\lambda)}] \sim \frac{(2\lambda)^{2n}}{(n!)^{2}}, \qquad \lambda \rightarrow \infty.
\end{equation}
In addition we determine the  asymptotics ($\lambda \rightarrow \infty$, fixed n) for the Fisher information of the Gegenbauer polynomials F[$\hat{C}_n^{(\lambda)}$] from (\ref{FG}), obtaining:
\begin{equation}
        \label{FC_lambda_infinity}
	F[\hat{C}_n^{(\lambda)}] =   (2 + 4 n)\lambda+ 2 + 4 n + 6 n^2 + \mathcal{O}(\lambda^{-1}),\qquad \lambda \rightarrow \infty.
\end{equation}

Finally, the substitution of the last two quantities into Eq. (\ref{fishershannonGeg}) gives rise to the following asymptotics for the Fisher-Shannon complexity of the orthonormal Gegenbauer polynomials:
\begin{equation}
        \label{CFSC_lambda_infinity}
	\mathcal{C}_{FS}[\hat{C}_n^{(\lambda)}] \sim  \frac{2^{4n}(2n+1)}{(n!)^{4}\pi e}\lambda^{4n+1} ,\qquad \lambda \rightarrow \infty.
\end{equation}
Note that the Fisher-Shannon complexity of the Gegenbauer polynomials $\hat{C}_n^{(\lambda)}(x)$ behaves dominantly according to the scaling law $\lambda^{4n+1}$ for large values of the parameter $\lambda$; so, very different to the Laguerre case (where this quantity is constant according to  (\ref{CFSLAG})). This is because the entropic Fisher and Shannon components behave according to laws $(\lambda,\lambda^{4n} )$ and $(\alpha^{-1}, \alpha)$ for the Gegenbauer and Laguerre cases with a given parameter $\lambda(\alpha)$, respectively. This indicates that when the orthogonality weight's parameter goes to infinity, the pointwise concentration around the polynomial nodes (as given by the Fisher information) linearly/inversely depends on the parameter in the Gegenbauer and Laguerre cases, respectively. And the disequilibrium/order of the Rakhmanov probability  follows a growth scaling law of $\lambda^{n}$ and $\alpha$ types for the Gegenbauer and Laguerre polynomials, respectively. \\

Finally, let us mention that expression (\ref{CFSC_lambda_infinity})  allows one to compute (a) the corresponding radial momentum complexity for the high-dimensional quantum states of the hydrogenic systems, because the radial eigenfunctions of such states are controlled by the Gegenbauer polynomials \cite{Yanez1994,Dehesa2001,Aptekarev2021} in momentum space, and (b) the corresponding angular momentum complexity for the high-dimensional quantum states of the hydrogenic and harmonic systems, because the angular eigenfunctions of such states are also controlled by the Gegenbauer polynomials.

\section{LMC complexity of the Gegenbauer polynomials}
\label{LMCGeg}

From expression (\ref{lmc}), the LMC complexity of the orthonormal Gegenbauer polynomials $\hat{C}_n^{(\lambda)}(x), \lambda> -\frac{1}{2}$, is given by
   \begin{equation}\label{lmcdefGeg}
      \mathcal{C}_{LMC}[\hat{C}_n^{(\lambda)}] = W_2[\hat{C}_n^{(\lambda)}] \times \mathcal{L}_{S}[\hat{C}_n^{(\lambda)}],
   \end{equation}
   where the second-order entropic moment $W_2[\hat{C}_n^{(\lambda)}]$ is, according to (\ref{renyinorm}), given by
   \begin{equation}\label{omega2Geg}
	W_2[\hat{C}_n^{(\lambda)}] = \int_{-1}^{+1}
\left(\left[\hat{C}_n^{(\lambda)}(x)\right]^2 h^{G}_{\lambda}(x)\right)^2\,dx = \int_{-1}^{+1} \,(1-x^2)^{2\lambda-1}\,\left[\hat{C}_n^{(\lambda)}(x)\right]^4\,dx.
\end{equation}
The explicit expression of this quantity at generic values of $n$ and $\lambda$ has not yet been determined in an analytically handy way, because neither $W_2[\hat{C}_n^{(\lambda)}]$ nor the spreading length $\mathcal{L}_{S}[\hat{C}_n^{(\lambda)}]$ are analytically known. In this section we obtain simple and compact analytical expressions for  $\mathcal{C}_{LMC}[\hat{C}_n^{(\lambda)}]$ in the two following extremal situations: when ($\lambda\rightarrow \infty; \mbox{fixed}\, n$) and when ($n\rightarrow \infty; \mbox{fixed}\, \lambda$). They are briefly summarized in Table  \ref{Table_G}.

 \subsection{Asymptotics $n\rightarrow \infty$}
 
 To obtain the LMC complexity $\mathcal{C}_{LMC}[\hat{C}_n^{(\lambda)}]$  in the limit ($n\rightarrow \infty; \mbox{fixed}\, \lambda$) we first realize that the asymptotical expression of $\mathcal{L}_{S}[\hat{C}_n^{(\lambda)}]$ has been already found in the previous section. To determine the remaining component, $W_2[\hat{C}_n^{(\lambda)}]$, when $n\rightarrow \infty$ we use Theorem 3 of Aptekarev et al \cite{Aptekarev2021}, obtaining
\begin{align}\label{omegaGeg_n_infinity}
	W_2[\hat{C}_n^{(\lambda)}] \sim
	\left\{
      \begin{array}{ll}
      n^{1-2\lambda},& -\frac{1}{2}<\lambda<\frac{1}{2}, \\ & \\
      \log{n},& \lambda=\frac{1}{2}, \\ & \\
     \frac{3}{2\pi^{3/2}}\frac{\Gamma(\lambda-\frac{1}{2})}{\Gamma(\lambda)}, & \lambda>\frac{1}{2}, 
%      \\ & \\
%      \infty, & {\rm otherwise}.
      \end{array}
      \right.
\end{align}
in the limit $n \rightarrow \infty $.

 This expression jointly with (\ref{ANLSG}) and (\ref{lmcdefGeg}) gives rise to the following asymptotical behavior ($n\rightarrow \infty$) of the LMC complexity of the orthonormal Gegenbauer polynomials $\hat{C}_n^{(\lambda)}(x)$:
 \begin{align}\label{lmcGeg_n_infinity}
      \mathcal{C}_{LMC}[\hat{C}_n^{(\lambda)}] \sim  \left\{
      \begin{array}{ll}
      \frac{\pi 2^{1-2\lambda}}{e}n^{1-2\lambda},& -\frac{1}{2}<\lambda<\frac{1}{2}, \\ & \\
     \frac{\pi}{e}\log{n},& \lambda=\frac{1}{2}, \\ & \\
     \frac{2^{-2\lambda}}{e}\frac{3}{\sqrt{\pi}}\frac{\Gamma(\lambda-\frac{1}{2})}{\Gamma(\lambda)}, & \lambda>\frac{1}{2}.
%      \\ & \\
%      \infty, & {\rm otherwise}.
      \end{array}
      \right.
   \end{align}
   
 Interestingly, the LMC complexity of the Gegenbauer polynomials follows a logarithmic growth scaling law (so, similarly to the Laguerre case (\ref{omega2inf})) at large degree $n$ only for $\lambda=\frac{1}{2}$. Nevetherless, this behavior has a qualitatively different origin.  Indeed, the two entropic components $(W_2,\mathcal{L}_{S})$ behave according to laws $(\log n, constant)$ and $(\frac{\log n}{n}, n)$ for the Gegenbauer and Laguerre cases, respectively. This indicates that when $n\rightarrow \infty$, the gradient content is much higher for the Gegenbauer polynomials  than for the Laguerre polynomials, while the disequilibrium (i.e., deviation from the uniform distribution) has the opposite behavior: in the Gegenbauer case it is much lower than in the Laguerre case for any fixed large degree. Moreover, note that the LMC complexity exponentially grows as $n^{1-2\lambda}$ for $-\frac{1}{2}<\lambda<\frac{1}{2}$ and has an uniform behavior (perfect disorder: non-dependence on $n$) for $\lambda>\frac{1}{2}$ when $n\rightarrow \infty$.

 Here again we remark that this mathematical result has relevant applications when we determine the spatial charge LMC complexity measures for the high-energy (Rydberg) states of hydrogenic and harmonic systems, and the total momentum LMC complexity measures for the high-energy (Rydberg) hydrogenic states.     \\

 \subsection{Asymptotics $\lambda\rightarrow \infty$}

 To determine the LMC complexity $\mathcal{C}_{LMC}[\hat{C}_n^{(\lambda)}]$  in the limit ($\lambda\rightarrow \infty; \mbox{fixed}\, n$) we first note that its first component, the spreading length $\mathcal{L}_{S}[\hat{C}_n^{(\lambda)}]$, has been already obtained in Eq. (\ref{LSCn}) above in subsection \ref{asympspreading}.\\
 
 Let us now tackle the second component, namely the second-order entropic moment $W_2[\hat{C}_n^{(\lambda)}]$ given by Eq. (\ref{omega2Geg}). We use the limiting relation (\ref{limitGeg}) into (\ref{omega2Geg}), obtaining for the orthogonal Gegenbauer polynomials the value
\begin{eqnarray}
	W_2[C_n^{(\lambda)}] &\sim
	& \left[C_n^{(\lambda)}(1)\right]^{4}\,\int_{-1}^{+1}(1-x^2)^{2\lambda-1}\,x^{4n}\,dx \nonumber\\
	 &=&\frac{(n+2\lambda-1)!^{4}}{n!^{4} \,(2\lambda-1)!^{4}}\frac{(1+(-1)^{4n})\Gamma(\frac{1}{2}+2n)\Gamma(2\lambda)}{2\Gamma(\frac{1}{2}+2n+2\lambda)}
\end{eqnarray}
Then, according to (\ref{omega2ortho}) one has the following asymptotics for the second-order entropic power of the orthonormal Gegenbauer polynomials:
\begin{equation}\label{W2oG}
	W_2[\hat{C}_n^{(\lambda)}] = \frac{1}{(\kappa_{n}^{G})^2}\,W_2[C_n^{(\lambda)}] = \frac{\Gamma(\frac{1}{2}+2n)}{\sqrt{2}\pi n!^{2}}\lambda^{\frac{1}{2}}+\mathcal{O}(\lambda^{-1/2}),\qquad \lambda\rightarrow \infty.
\end{equation}
Finally, the combination of expressions (\ref{lmcdefGeg}), (\ref{LSCn}) and (\ref{W2oG}) lead to theasymptotical behavior
\begin{equation}\label{lmcGeg_lambda_infinity}
      \mathcal{C}_{LMC}[\hat{C}_n^{(\lambda)}] =  \frac{2^{\frac{n-1}{2}}\Gamma(\frac{1}{2}+2n)}{\pi n!^{\frac{5}{2}}}\lambda^{\frac{n+1}{2}}+\mathcal{O}(\lambda^{\frac{n}{2}}). \qquad \lambda \rightarrow \infty 
   \end{equation}
   for the LMC complexity of the (orthonormal) Gegenbauer polynomials. Note that the LMC complexity of the Gegenbauer polynomials $\hat{C}_n^{(\lambda)}(x)$ behaves dominantly according to the scaling law $\lambda^{(n+1)/2}$ for large values of the parameter $\lambda$; so, different to the Laguerre polynomials $\hat{L}_n^{(\alpha)}(x)$  (where this quantity behaves as $\alpha^{2n}$; see (\ref{CLMCLag})). This is because the two entropic components $(W_2,\mathcal{L}_{S})$ behave according to laws $(\lambda^{1/2},\lambda^{n/2} )$ and $(\alpha^{2n-1/2}, \alpha^{1/2})$ for the Gegenbauer and Laguerre cases with a given polynomial degree, respectively. This indicates that when the orthogonality weight's parameter goes to infinity, the disequilibrium/order (as given by the second-order entropic moment) depends on the parameter as $\lambda^{1/2}$ and $\alpha^{2n-1/2}$ in the Gegenbauer and Laguerre cases, respectively. And the disorder of the Rakhmanov probability (as given by the Shannon entropy power) follows a growth scaling law of $\lambda^{n/2}$ and uniform types for Gegenbauer and Laguerre polynomials, respectively.  
   Finally, let us remark that this mathematical result has relevant applications when we determine the spatial charge LMC complexity measures for the high-dimensional (quasi-classical) states of hydrogenic and harmonic systems, and the total momentum LMC complexity measures for the high-dimensional hydrogenic states.

\section{Conclusions} \label{Conclud}

In this work we investigate the notions of simplicity/complexity and order/disorder for the parameter-dependent hypergeometric orthogonal polynomials of Laguerre and Gegenbauer types. This is done by means of the Fisher-Shannon and LMC complexity measures of the associated Rakhmanov probability density of such polynomials. Each of these quantities capture two configurational facets of the HOPs: the Shannon spreading length or entropy power of the polynomials (which quantifies the equilibrium/disorder of the Rakhmanov probability) and the deviation from equilibrium or disequilibrium/order (which is measured by the Fisher information and the second-order entropic moment in the Fisher-Shannon and LMC complexity measures, respectively).\\

We have determined the Fisher-Shannon and LMC complexities of the Laguerre and Gegenbauer polynomials in the two following asymptotics at first order: when ($n\rightarrow \infty; \mbox{fixed polynomial's parameter}$) and when ($\text{parameter}\rightarrow \infty; \mbox{fixed}\, n$). We have found the following results. First, in the aymptotics ($n\rightarrow \infty; \mbox{fixed parameter}$) the Fisher-Shannon measure of both Laguerre and Gegenbauer polynomials follow a simple exponential power ($n^3$)-law. However, the LMC complexities of these two sets of polynomials with high degree have a similar logarithmic behavior only for the Gegenbauer parameter $\lambda = 1/2$, while the LMC measure of the Gegenbauer polynomials follows an exponential and constancy (i.e., it does not depend on $n$) behavior for $\lambda < 1/2$, and $> 1/2$, respectively.

Second, in the asymptotics ($\alpha\rightarrow \infty; \mbox{fixed}\, n$)  the Fisher-Shannon measure of Laguerre polynomials $L_n^{(\alpha)}(x)$  gets constancy (i.e., it does not depend on $\alpha$), while the LMC measure of such polynomials follow the power law $\alpha^n$. Moreover, something similar happens for the Gegenbauer polynomials $\hat{C}_n^{(\lambda)}(x)$  when ($\lambda\rightarrow \infty; \mbox{fixed}\, n$); namely, the Fisher-Shannon and LMC measures behave according to the power laws $\lambda^{n+1}$ and $\lambda^{(n+1)/2}$, respectively.\\

 Finally, these different scaling laws can be understood by observing the contributions of the two entropic components of the complexity measures in each case. Particularly interesting is the constancy of the Fisher-Shannon complexity of the Laguerre polynomials $L_n^{(\alpha)}(x)$ when ($\alpha\rightarrow \infty; \mbox{fixed}\, n$); this is because the $\alpha$-dependence of  Shannon and Fisher components of this measure mutually cancel, indicating uniformity (so, perfect disorder) since the Fisher-Shannon complexity does not depend on $\alpha$.\\
 
  These mathematical results are interesting \textit{per se} and because of their applications to compute the physical entropy and complexity measures of the charge and momentum distributions of the high-dimensional (quasi-classical) and high-energy (Rydberg) quantum states of the multidimensional atomic systems, such as e.g the Coulomb and harmonic systems as previously pointed out. This should not be surprising because the charge (momentum) probability density of (e.g.) multidimensional hydrogenic and harmonic oscillator systems can be represented by the Rakhmanov density of the Laguerre and Gegenbauer polynomials in position (momentum) space, respectively.

\section*{Acknowledgments}
This work has been partially supported by the Agencia Estatal de Investigaci\'on (Spain) and the European Regional Development Fund (FEDER) under the grant PID2020-113390GB-I00.

\end{document}